# The Dirac Equation from a Bohmian Point of View

Sergio Hernández-Zapata[1,2]

[1] Facultad de Ciencias, Universidad Nacional Autónoma de México, Circuito exterior de Ciudad Universitaria, 04510 Distrito Federal, México.

[2] Author address: shz@fciencias.unam.mx

## *Abstract*

In this paper we intend to extend some ideas of a recently proposed Lorentz-invariant Bohmian model, obeying Klein-Gordon equations, but considering particles with a spin different than zero. First we build a Bohmian model for a single particle, closely related to the Bohm-Dirac model, but using Nikolic's ideas on a space-time probability density and the introduction of a synchronization parameter $\sigma$ for the particle trajectory. We prove that, in the case of a spinor whose component phases become equal when $\hbar \to 0$, it is possible to use the usual Bohmian Mechanics procedures to obtain the classical limit. We extend these ideas for a two-particle system, based on two non-local but covariant Dirac-like equations.

### 1. Introduction

From the Bohmian point of view there is no need for big changes to include, on the theory, particles with spin different than zero. In spite of some conceptions about spin, including intuitions by David Bohm [1], in further developments of the Bohmian Mechanics it is not considered that particles are intrinsically spinning (see Reference [2] for a very good introduction to Bohmian Mechanics in a general way including spin). The usual analogy with a little planet spinning around its own axis has been fully abandoned by the modern Bohmian theory.

Bell says that the only hidden variables to be considered are the particle positions. He states explicitly that this is a great merit on Bohm's theory. His reasoning is that every measure on a laboratory can be reduced to positions. For example, the position of the needle on a measurement apparatus or the electron positions when hitting an oscilloscope screen. It is somewhat odd, according to Bell, to consider the position of the particles as a hidden variable. Precisely, it is the position what immediately reveals itself on a Quantum Mechanics experiment. From Bell's point of view, position would be the measured variable and the wave function would be "the Revealed Variable". The wave character shows up after doing experiments with a big number of particles [3].

However, the term "hidden variable" is important in the history of Quantum Mechanics. It is usually considered under that term anything being added to the wave function with the purpose, for example, of building an Objective Quantum Interpretation, or also, though is not the same, recover the deterministic character of Physics on this level. It is from that point of view and only from that point of view, that the Bohm Theory is a Hidden Variable Theory. The spin, although this term may sound anachronistic [4,5], would be a property of the wave function and not one of the particle. Therefore, it cannot be considered another hidden variable. In a Stern-Gerlach experiment where you put a magnetic field for the particle to pass through, depending on *the position where the particle is finally detected* we say we obtain as a result "up" or "down" [4,5].

In Non-Relativistic Bohmian Mechanics a spin-1/2 particle wave function can be thought as a function:

$$\varphi \colon \mathbb{R}^3 \quad \to \quad \mathbb{C}^2,$$

the so called spinor. Or, in a different notation,

$$\begin{pmatrix} \varphi_1 \\ \varphi_2 \end{pmatrix} = \begin{pmatrix} \varphi_1(\vec{x}) \\ \varphi_2(\vec{x}) \end{pmatrix},$$

where $\vec{x}$ is called the generic position of the particle. We will leave the time dependence implicit. That is, all the wave functions considered, in general, depend also on time. So on this introduction instead of writing $\varphi_t$ as in [4], we will simply write $\varphi$.

Now, for *N* particles the wave function could be thought as a function:

$$\varphi \colon \mathbb{R}^{3N} \quad \to \quad \mathbb{C}^{2^N}.$$

On this introduction we will restrict the discussion to two particles for clarity purposes. Then

$$\varphi \colon \mathbb{R}^6 \quad \to \quad \mathbb{C}^4$$

or

$$\varphi = \begin{pmatrix} \varphi_{11}\left(\vec{x}^{(1)}, \vec{x}^{(2)}\right) & \varphi_{12}\left(\vec{x}^{(1)}, \vec{x}^{(2)}\right) \\ \varphi_{21}\left(\vec{x}^{(1)}, \vec{x}^{(2)}\right) & \varphi_{22}\left(\vec{x}^{(1)}, \vec{x}^{(2)}\right) \end{pmatrix}.$$

Here $\vec{x}^{(1)}, \vec{x}^{(2)}$ are the two particles generic coordinates. The (Born) probability distribution for the two particles is given by

$$\varphi_{il}^* \delta_{ij}\, \delta_{lm}\, \varphi_{jm} = |\varphi_{11}|^2 + |\varphi_{12}|^2 + |\varphi_{21}|^2 + |\varphi_{22}|^2. \qquad (1.1)$$

In the left-hand side term we have summed on the repeated matrix indexes. The particle density currents are determined by

$$\begin{aligned} \vec{J}^{(1)} &= \frac{\hbar}{2im_1}\left\{\varphi_{il}^* \delta_{ij}\, \delta_{lm}\, \nabla^{(1)} \varphi_{jm} - \varphi_{il}\, \delta_{ij}\, \delta_{lm}\, \nabla^{(1)} \varphi_{jm}^*\right\} \\ \vec{J}^{(2)} &= \frac{\hbar}{2im_2}\left\{\varphi_{il}^* \delta_{ij}\, \delta_{lm}\, \nabla^{(2)} \varphi_{jm} - \varphi_{il}\, \delta_{ij}\, \delta_{lm}\, \nabla^{(2)} \varphi_{jm}^*\right\}. \end{aligned} \qquad (1.2)$$

On the base of (1.1) and (1.2) we can propose the Bohm-like guide equations for the particles in the standard way:

$$\frac{d\vec{X}^{(1)}}{dt} = \frac{\hbar}{2im_1}\frac{\varphi_{il}^{*}\delta_{ij}\,\delta_{lm}\nabla^{(1)}\varphi_{jm} - \varphi_{il}\delta_{ij}\,\delta_{lm}\nabla^{(1)}\varphi_{jm}^{*}}{\varphi_{il}^{*}\delta_{ij}\,\delta_{lm}\,\varphi_{jm}}$$
$$\frac{d\vec{X}^{(2)}}{dt} = \frac{\hbar}{2im_2}\frac{\varphi_{il}^{*}\delta_{ij}\,\delta_{lm}\nabla^{(2)}\varphi_{jm} - \varphi_{il}\delta_{ij}\,\delta_{lm}\nabla^{(2)}\varphi_{jm}^{*}}{\varphi_{il}^{*}\delta_{ij}\,\delta_{lm}\,\varphi_{jm}}. \qquad (1.3)$$

In these expressions we sum on repeated indexes independently on the numerator and denominator. Also on the right-hand side, both expressions are evaluated on the particles configurational or actual coordinates $\vec{X}^{(1)}$, $\vec{X}^{(2)}$. Then we prove, in a standard way that if the particles are distributed according to (1.1) at time *t*=0, once this expression has been normalized, it will remain distributed that way for every time *t*. These are very well known results of Non-Relativistic Bohmian Mechanics.

Let us suppose, as Bell does [3] the particular case when $\varphi_{il}$ is separable that is $\varphi_{il}\left(\vec{x}^{(1)}, \vec{x}^{(2)}\right) = \chi_i\left(\vec{x}^{(1)}\right)\xi_l\left(\vec{x}^{(2)}\right)$. Then

$$\frac{d\vec{X}^{(1)}}{dt} = \frac{\hbar}{2im_1}\frac{\chi_i^{*}\xi_l^{*}\delta_{ij}\,\delta_{lm}\,\xi_m\nabla^{(1)}\chi_j - \chi_i\xi_l\delta_{ij}\,\delta_{lm}\,\xi_m^{*}\nabla^{(1)}\chi_j^{*}}{\chi_i^{*}\xi_l^{*}\delta_{ij}\,\delta_{lm}\,\chi_j\,\xi_m}$$
$$= \frac{\hbar}{2im_1}\frac{(\xi_l^{*}\delta_{lm}\,\xi_m)\chi_i^{*}\delta_{ij}\,\nabla^{(1)}\chi_j - (\xi_l\delta_{lm}\,\xi_m^{*})\chi_i\delta_{ij}\,\nabla^{(1)}\chi_j^{*}}{\left(\chi_i^{*}\delta_{ij}\,\chi_j\right)(\xi_l^{*}\delta_{lm}\,\xi_m)}.$$

Now $\xi_l^{*}\delta_{lm}\,\xi_m = \xi_l\delta_{lm}\,\xi_m^{*}$ since it is real. Therefore

$$\frac{d\vec{X}^{(1)}}{dt} = \frac{\hbar}{2im_1}\frac{\chi_i^{*}\delta_{ij}\,\nabla^{(1)}\chi_j - \chi_i\delta_{ij}\,\nabla^{(1)}\chi_j^{*}}{\chi_i^{*}\delta_{ij}\,\chi_j} = \frac{\hbar}{m_1}\frac{Im\left(\chi^{+}\nabla^{(1)}\chi\right)}{\chi^{+}\chi}$$

and analogously

$$\frac{d\vec{X}^{(1)}}{dt} = \frac{\hbar}{m_2}\frac{Im\left(\xi^{+}\nabla^{(2)}\xi\right)}{\xi^{+}\xi}.$$

This shows the difference between a separable wave function and a not-separable one in Non-Relativistic Bohmian Mechanics, following Bell's ideas [3].

In Sect. 2 of this paper we extend this kind of ideas to Special Relativity. We base the study on the Dirac equation, the relativistic equation for the electron. We use Minkowski notation for Special Relativity in which it is not needed to distinguish between covariant and contravariant indexes [6]. On this notation the coordinates of an event are denoted by $\vec{x}, ict$. In this case the wave function is a four component spinor. We also assume that the particle is interacting with an external electromagnetic field, measured in Gaussian units. As it is usual in Bohmian literature, we supplement the wave function with a Bohm-like equation that establish in a precise way how the particle moves. We want the Bohmian velocities and the whole model to be covariant. We then make a proposal for a guide equation considering the following ideas by Nikolic [7]: In the first place we consider an space-time probability density instead of an spatial density. We introduce an equivariance (quantum equilibrium) condition by introducing a scalar parameter $\sigma$ with time units [7,8]. We assume that the probability distribution at time $\sigma = 0$ is given by our space-time density. From this it is proven that the particle probability density will remain so for every $\sigma$. Since the wave function, the spinor, does not depend on $\sigma$, we have a stationary flow with respect to this parameter with a space-time distribution that is also a scalar [7].

The only way to prove, for now, that the model presented here is physically reasonable is to study the classical limit [9]. We do this on Sect. 3. We do not consider the classical limit in general, but just in a case we think it is particularly important. Under a certain hypothesis on spinor, that its components phases being equal on the classical limit. We can prove on this hypothesis that: *(a)* The interval $d\sigma$ tends to the proper time of the particle $d\tau$. This is analogous to the study of the classical limit in [10] for a *N*-particle model based on the Klein-Gordon equation [7,10]. *(b)* The classical movement law of the particle obtained is the one for classical Einsteinian Mechanics for one particle subject to the Lorentz Force. *(c)* Two kinds of classical solutions are obtained in different regions of space-time. One with a negative charge *e*, the electron charge, the other one with a positive charge *–e.*

We believe that the hypothesis of phase equality on the classical limit is not entirely arbitrary. This hypothesis tells us that the resulting classical system has a well defined action. Let us suppose a fixed action $S$ (function of the space-time coordinates) that fits as the classical limit of the component phases of a set of spinors. It can be proved that this set is a subspace of the solution space of spinors from Dirac equation. We conjecture that the Bohmian problem based on Dirac equation, restricted to this subspace, is particularly physically relevant. We believe that a very similar hypothesis can be used for the classical limit following the method of [9] on the Non-Relativistic version of our model, Pauli equation from a Bohmian point of view.

In Sect. 4 we study the model extension for two particles. The particles interact with an external electromagnetic field but there is no electromagnetic interaction with each other. It is possible to build a model for N particles (Some kind of ideal gas for particles with spin different than zero [7, 10]). For the sake of simplicity we will work with only two particles. We use the Nikolic's idea [7, 10] of a density on the space-time configuration space of the two particles $\vec{x}^{(1)}, ict^{(1)}$ , $\vec{x}^{(2)}$, $ict^{(2)}$. As it can be seen the model is multitemporal. The times $ict^{(1)}$, $ict^{(2)}$ are treated on an equal footing with spatial generic variables. The Wave Function, a 16-component spinor, is supplemented with Bohm-like equations. These equations determine how the configurational variables $\vec{X}^{(1)}, icT^{(1)}$ , $\vec{X}^{(2)}$, $icT^{(2)}$ evolve in terms of a parameter $\sigma$. This parameter, with time units, is a scalar and represents an extension of [7] to a model built on Dirac equation. *The model is Non-Local and at the same time Covariant and some ideas of Bell* [3] *are analyzed on the Special Relativity context.*

Finally, in Sect. 5, we present our conclusions.

**2. A Modification of The Bohm-Dirac Model.**

In this paper we will develop a Bohm-like theory for the Dirac wave equation in the one particle case. We start with the Dirac equation:

$$\gamma_\mu \left(p_\mu - \frac{e}{c} A_\mu\right) \psi = -mc\psi \qquad (2.1)$$

where

$$\psi = \begin{pmatrix} \psi_1 \\ \psi_2 \\ \psi_3 \\ \psi_4 \end{pmatrix},$$

the spinor, plays the role of the wave function for one particle [11-12]. On the other hand, $A_\mu$ represents an external electromagnetic field, evaluated on the particle generic coordinates, $e$ is a negative charge equal to that of the electron, and the matrixes $\gamma_\mu$ are defined in terms of the 2x2 Pauli matrixes as follows:

$$\gamma_i = \begin{pmatrix} 0 & \sigma_i \\ -\sigma_i & 0 \end{pmatrix} ; i = 1, 2, 3$$

where

$$\sigma_1 = \begin{pmatrix} 0 & 1 \\ 1 & 0 \end{pmatrix}, \qquad \sigma_2 = \begin{pmatrix} 0 & -i \\ i & 0 \end{pmatrix}, \qquad \sigma_3 = \begin{pmatrix} 1 & 0 \\ 0 & -1 \end{pmatrix}$$

and $\gamma_4 = i\beta$ where $\beta = \begin{pmatrix} 1 & 0 & 0 & 0 \\ 0 & 1 & 0 & 0 \\ 0 & 0 & -1 & 0 \\ 0 & 0 & 0 & -1 \end{pmatrix}$.

Usually it is defined $\bar{\psi} = \psi^+ \beta$ where $\psi^+ = ({\psi_1}^* \;\; {\psi_2}^* \;\; {\psi_3}^* \;\; {\psi_4}^*)$.

In this case

$$\bar{\psi}\psi = |\psi_1|^2 + |\psi_2|^2 - |\psi_3|^2 - |\psi_4|^2$$

is real [11-12]. The expression (our candidate for space-time probability density)

$$|\bar{\psi}\psi| \qquad (2.2)$$

is a scalar. We use absolute value because we want a positive definite expression for the space-time probability density. The term $\bar{\psi}\psi$ could easily be negative on certain regions of space-time. In the first place $\psi' = S\psi$, where $S$ is the transformation matrix that transforms the spinor from one frame of reference to another. Therefore $\psi'^+ = \psi^+ S^+$ and $\psi'^+ \beta = \psi^+ S^+ \beta = \psi^+ \beta S^{-1}$. Here we have used the identity $S^+ \beta = \beta S^{-1}$. By multiplying both sides of the identity by $\psi' = S\psi$ we have $\psi^{'+}\beta\psi^{'} = \psi^{+}\beta S^{-1}S\psi = \psi^{+}\beta\psi$. Or $\overline{\psi'}\psi' = \bar{\psi}\psi$ and considering the absolute value on both sides we obtain that $\left|\overline{\psi'}\psi'\right| = \left|\bar{\psi}\psi\right|$. This tells us (2.2) is in fact a scalar.

The following are very well known facts. If $\psi$ satisfies the Dirac equation then

$$c\bar{\psi}\gamma_{\mu}\psi \qquad (2.3)$$

is a four-vector and

$$\frac{\partial}{\partial X_{\mu}}\left(c\bar{\psi}\gamma_{\mu}\psi\right) = 0. \qquad (2.4)$$

From a Bohmian point of view Equation (2.1) does not completely describe the dynamics of the particle. We have to supplement (2.1) with a Bohm-like equation that describes how the velocity of the particle is determined by the spinor $\psi$. We propose

$$V_{\mu} = \frac{dX_{\mu}}{d\sigma} = \frac{c\bar{\psi}\gamma_{\mu}\psi}{\left|\bar{\psi}\psi\right|} \qquad (2.5)$$

where the right-hand side of the equation is evaluated on the particle configurational coordinates $\vec{X}, icT$. On the other hand, $d\sigma$ is a parameter with time units that we will use to describe the dynamics (Let us call $d\sigma$ "Nikolic time for the model", since we are using a very similar idea to the one used by Nikolic in [7] to build a Lorentz-Invariant Bohmian Model, based on the Klein-Gordon equation). $d\sigma$ is clearly a scalar since $dX_{\mu}$ and $c\bar{\psi}\gamma_{\mu}\psi/\left|\bar{\psi}\psi\right|$ are four-vectors. Let's rewrite (2.4) as

$$\frac{\partial}{\partial X_{\mu}}\left(\left|\bar{\psi}\psi\right|V_{\mu}\right) = 0. \qquad (2.6)$$

On the same way than in Non-Relativistic Bohmian Mechanics, the system dynamics must be supplemented with a probabilistic hypothesis on initial conditions. Let us assume that the probability that the particle position four-vector $X_{\mu}$ is in the space-time volume $\Delta x\Delta y\Delta zc\Delta t$ surrounding $x_{\mu}$ on $\sigma = 0$ is proportional to $\rho(x_{\mu})\Delta x\Delta y\Delta zc\Delta t = \left|\bar{\psi}\psi\right|\Delta x\Delta y\Delta zc\Delta t$. Thus, as a consequence of the equations (2.5) and (2.6), the particle position four-vector will continue to be distributed according to the same probability density [7, 10]. Since we are using a very similar idea to Nikolic's [7] in the construction of the model, we will call $\left|\bar{\psi}\psi\right|$, once normalized, *Born-Nikolic probability distribution for the model.*

In Special Relativity the volume of a region of space-time is independent on the frame of reference used for its calculation. This condition is given by

$$\frac{\partial(x^{'}, y^{'}, z^{'}, ct^{'})}{\partial(x, y, z, ct)} = 1.$$

It follows that the volume of a region of space-time is also Lorentz-invariant [10]. The probability assigned by the Born-Nikolic rule (for some value of $\sigma$) to a given region V of the space-time for the particle is given by the integral of $\rho(x_\mu) = |\bar{\psi}\psi|$ on V. It follows that the probability assigned to a certain region V in an inertial frame of reference will be the same than the probability assigned to the Lorentz-transformed region V' in a different frame of reference. In other words, the calculation of the probability is independent on the frame of reference used. Therefore, if the particle position four-vector is distributed according to the Born-Nikolic rule proposed here in an inertial frame of reference for some value of $\sigma$ *they will be so distributed for all inertial frames of reference and for all values of* $\sigma$. *The "quantum equilibrium condition" is thus, in the model, Lorentz-invariant* [7, 10].

A very important hypothesis that we are considering, not so much in the construction of the model, but in the probability considerations is that $\rho(x_\mu) = |\bar{\psi}\psi|$ is normalisable. This is not proved in any way. Nevertheless, let us consider for a moment, as an exercise, $\psi^+\psi$ as a space-time probability density (This disagrees with Nikolic's proposal [7, 10] because this quantity is not scalar but the fourth component of a four-vector). Now, the integral of $\psi^+\psi$ with respect to space is time-independent due to the Dirac continuity equation (2.4). If after that we integrate respect to time the integral diverges and so $\psi^+\psi$ is not normalisable if we think of it as a space-time density. However this argument cannot be applied to $|\bar{\psi}\psi|$ because the spatial integral of this quantity is not constant on time. We will consider the provisional hypothesis that there is no problem with normalization of this function in space-time.

In the following we consider the relations $\beta\gamma_i = \alpha_i; i = 1,2,3$ and $\beta\gamma_4 = i$. It is important to compare the exposed ideas and some important ideas of Bohm's school about construction of a Bohmian Model for Dirac equation. Our equation (2.5) can be decomposed on spatial and temporal parts as follows

$$\frac{dX_i}{d\sigma} = \frac{c\bar{\psi}\gamma_i\psi}{|\bar{\psi}\psi|} = \frac{c\psi^+\beta\gamma_i\psi}{|\bar{\psi}\psi|} = \frac{c\psi^+\alpha_i\psi}{|\bar{\psi}\psi|}; \quad \frac{dicT}{d\sigma} = \frac{c\bar{\psi}\gamma_4\psi}{|\bar{\psi}\psi|} = \frac{c\psi^+\beta\gamma_4\psi}{|\bar{\psi}\psi|} = \frac{ic\psi^+\beta^2\psi}{|\bar{\psi}\psi|} = \frac{ic\psi^+\psi}{|\bar{\psi}\psi|}.$$

It follows that $dT/d\sigma = \psi^+\psi/|\bar{\psi}\psi|$ which implies

$$\frac{d\vec{X}}{dT} = \frac{c\psi^+\vec{\alpha}\,\psi}{\psi^+\psi}. \qquad (2.7)$$

This constitutes the Bohm-Dirac theory as formulated by David Bohm [13-14]. We can consider the right-hand side of (2.7) as a function of the configurational coordinates $\vec{X}$ and time $T$. Then the wave equation can be written as

$$i\hbar\frac{\partial\psi}{\partial T} = c\vec{\alpha}\cdot\left(\vec{p} - \frac{e}{c}\vec{A}\right)\psi + (e\phi + mc^2\beta)\psi.$$

Here the spinor is seen as a function of the generic coordinates $\vec{x}$ of the particle and configurational time $T$. The continuity equation can be written as

$$\frac{\partial}{\partial x_i}(c\psi^+\alpha_i\psi)+\frac{\partial}{\partial T}(\psi^+\psi)=0. \qquad (2.8)$$

Then if at time $T=0$ the particle is distributed with probability $\psi^+\psi$, once normalized the function in the space, this will remain so for every $T$. This follows from the continuity equation (2.8) and the Bohm-Dirac equation (2.7). This property of "quantum equilibrium" is only valid on a preferential frame of reference. That is, *there's quantum equilibrium in a preferential frame of reference but not in any frame of reference.*

Since what we are interested, like Nikolic [7, 10], in a space-time density in the relativistic context we will work with the guide equation (2.5) and the space-time density $\left|\bar{\psi}\psi\right|$ rather than with the guide equation (2.7) and the space density $\psi^+\psi$. Our idea is that this procedure (insisting on the space-time probability density) allows us to extend the model to a Non-Local model for *N* particles that remains Lorentz-Invariant.

## 3. Classical Limit

We now intend to extend the ideas of [9] to obtain the classical limit. We will not do it in general but only for spinors that satisfy the following hypothesis.

*Hypothesis of phases equality on the classical limit.* Let $\psi_i=R_i exp(i\,S_i/\hbar)$ be the $i$-th component of the spinor. Let us assume that

$$\lim_{\hbar\to 0} S_i=S$$

where $S$ is independent on $i$ $(i=1,2,3,4)$. In this section and only in this section we will use $S$ for that limit. On the limit when $\hbar$ tends to 0 the phases of the different components of the spinor become equal to each other. If we write $S_i$ as a series in $\hbar$ we obtain

$$S_i=S+S_i^{(1)}\hbar+S_i^{(2)}\hbar^2+S_i^{(3)}\hbar^3+\cdots$$

where $S$ is independent on the index $i$. Observation: $\left[S_i^{(n)}\hbar^n\right]=[\hbar]$, that is, every term on the series has units of action. Always that the spinor has the form just described we will say that we are using *the hypothesis of phases equality*. Let us consider the set of spinors with the property that the hypothesis of phases equality is satisfied on the limit when $\hbar\to 0$ in such a way that $S$ is a fixed function. It can be proved that this set is a subspace of the space of all spinors that satisfy the Dirac equation. We may consider a Bohmian dynamics restricted to this subspace of spinors. As a clarification, if we add two spinors in different subspaces of this kind the sum spinor does not satisfy the hypothesis of phases equality on the classical limit.

For calculation convenience, let us define $\psi^c=e^{-i\frac{S}{\hbar}}\psi$. From the hypothesis of phases equality on the classical limit we can define the "classical spinor" as

$$\psi^{class} = \lim_{\hbar\to 0}\psi^c = \lim_{\hbar\to 0} e^{-i\frac{S}{\hbar}}\psi.$$

The "classical spinor" is only correctly defined when using a hypothesis as pointed out before. This spinor does not depend on $\hbar$ and its components are $\psi_i^{class} = R_i exp\left(iS_i^{(1)}\right)$.

Now, if $\psi$ satisfies the Dirac equation then $\psi$ satisfies the equation

$$\left(p_\mu - \frac{e}{c}A_\mu\right)\left(p_\mu - \frac{e}{c}A_\mu\right)\psi + \frac{ie\hbar}{2c}\gamma_\mu\gamma_\nu F_{\mu\nu}\psi = -m^2c^2\psi. \qquad (3.1)$$

The reciprocal is not true. Equation (3.1) has more solutions than the Dirac equation [11]. Equation (3.1) is very similar to the Klein-Gordon equation except for an additional term that is in a certain way the relativistic extension of the Stern-Gerlach term of the Pauli equation. Let us rewrite Equation (3.1) making explicit every operator:

$$-\hbar^2\frac{\partial^2\psi}{\partial x_\mu\partial x_\mu} - \frac{e\hbar}{ic}\frac{\partial A_\mu}{\partial x_\mu}\psi - \frac{2e\hbar}{ic}A_\mu\frac{\partial\psi}{\partial x_\mu} + \frac{e^2}{c^2}A_\mu A_\mu\psi + \frac{ie\hbar}{2c}\gamma_\mu\gamma_\nu F_{\mu\nu}\psi = -m^2c^2\psi.$$

Let us consider the following facts, always under the hypothesis of the phases equality on the classical limit (Appendix 1):

i) $\lim_{\hbar\to 0}\left(-\hbar^2 e^{-i\frac{S}{\hbar}}\frac{\partial^2\psi}{\partial x_\mu\partial x_\mu}\right) = \frac{\partial S}{\partial x_\mu}\frac{\partial S}{\partial x_\mu}\psi^{class}$

ii) $\lim_{\hbar\to 0}\left(-\frac{e\hbar}{ic}e^{-i\frac{S}{\hbar}}\frac{\partial A_\mu}{\partial x_\mu}\psi\right) = 0$

iii) $\lim_{\hbar\to 0}\left(-\frac{2e\hbar}{ic}e^{-i\frac{S}{\hbar}}A_\mu\frac{\partial\psi}{\partial x_\mu}\right) = -\frac{2e}{c}\frac{\partial S}{\partial x_\mu}A_\mu\psi^{class}$

iv) $\lim_{\hbar\to 0}\left(\frac{e^2}{c^2}A_\mu A_\mu e^{-i\frac{S}{\hbar}}\psi\right) = \frac{e^2}{c^2}A_\mu A_\mu\psi^{class}$

v) $\lim_{\hbar\to 0}\left(\frac{ie\hbar}{2c}\gamma_\mu\gamma_\nu F_{\mu\nu}e^{-i\frac{S}{\hbar}}\psi\right) = 0$

vi) $\lim_{\hbar\to 0}\left(-m^2c^2e^{-i\frac{S}{\hbar}}\psi\right) = -m^2c^2\psi^{class}$.

Therefore

$$\frac{\partial S}{\partial x_\mu}\frac{\partial S}{\partial x_\mu}\psi^{class} - \frac{2e}{c}\frac{\partial S}{\partial x_\mu}A_\mu\psi^{class} + \frac{e^2}{c^2}A_\mu A_\mu\psi^{class} = -m^2c^2\psi^{class}$$

Given that we suppose $\psi^{class} \neq 0$, after some rewriting we have:

$$\left(\frac{\partial S}{\partial x_\mu} - \frac{e}{c}A_\mu\right)\left(\frac{\partial S}{\partial x_\mu} - \frac{e}{c}A_\mu\right) = -m^2c^2. \qquad (3.2)$$

Which tells us that $S$ satisfies the so called Hamilton-Jacobi Relativistic Equation. Thus the first step of our argument is complete.

Let us study the guide equation (2.5). Rewriting the Dirac equation on the following way:

$$\gamma_\mu\left(\frac{\hbar}{i}\frac{\partial}{\partial x_\mu}-\frac{e}{c}A_\mu\right)\left(e^{i\frac{S}{\hbar}}\psi^c\right)=-mce^{i\frac{S}{\hbar}}\psi^c$$

$$\gamma_\mu\left(\frac{\hbar}{i}e^{i\frac{S}{\hbar}}\frac{\partial\psi^c}{\partial x_\mu}+\frac{\partial S}{\partial x_\mu}e^{i\frac{S}{\hbar}}\psi^c-\frac{e}{c}A_\mu e^{i\frac{S}{\hbar}}\psi^c\right)=-mce^{i\frac{S}{\hbar}}\psi^c.$$

The limit of the last expression multiplied by $e^{-i\frac{S}{\hbar}}$ when $\hbar\to 0$ is

$$\gamma_\mu\left(\frac{1}{m}\frac{\partial S}{\partial x_\mu}-\frac{e}{mc}A_\mu\right)\psi^{class}=-c\psi^{class}. \qquad (3.3)$$

Let's use the following notation $U_\mu=\frac{1}{m}\frac{\partial S}{\partial x_\mu}-\frac{e}{mc}A_\mu$.

Then (3.3) can be written as a matrix equation:

$$\begin{pmatrix} iU_4+c & 0 & U_3 & U_1-iU_2 \\ 0 & iU_4+c & U_1+iU_2 & -U_3 \\ -U_3 & -U_1+iU_2 & -iU_4+c & 0 \\ -U_1-iU_2 & U_3 & 0 & -iU_4+c \end{pmatrix}\begin{pmatrix}\psi_1^{class}\\ \psi_2^{class}\\ \psi_3^{class}\\ \psi_4^{class}\end{pmatrix}=\begin{pmatrix}0\\0\\0\\0\end{pmatrix}.$$

This equation has non-trivial solutions since the determinant is zero. The solution space is a two dimension space, so we write the components of the classical spinor in terms of $\psi_1^{class}$, $\psi_2^{class}$

$$\psi_1^{class}=R_1e^{i\frac{S}{\hbar}},\qquad \psi_2^{class}=R_2e^{i\frac{S}{\hbar}},\quad \psi_3^{class}=\frac{U_3\psi_1^{class}+(U_1-iU_2)\psi_2^{class}}{-iU_4+c},$$

$$\psi_4^{class}=\frac{(U_1+iU_2)\psi_1^{class}-U_3\psi_2^{class}}{-iU_4+c}.$$

So we can build the analogous of $\bar{\psi}\psi$ but using the "classical spinor"

$$\overline{\psi^{class}}\psi^{class}=\frac{(2c^2-2icU_4)(R_1^2+R_2^2)}{-U_4^2-2icU_4+c^2}=\frac{2ic(R_1^2+R_2^2)}{(U_4+ic)}. \qquad (3.4)$$

Since $U_4=i\left(-\frac{1}{mc}\frac{\partial S}{\partial t}-\frac{e}{mc}\phi\right)$ is purely imaginary $\overline{\psi^{class}}\psi^{class}$ is real. This should be so because we proved that $\bar{\psi}\psi$ is real. Now, the classical version of the current $c\bar{\psi}\gamma_\mu\psi$ that can be calculated from the expressions for the "classical spinor" components is

$$c\overline{\psi^{class}}\gamma_\mu\psi^{class}=\frac{2icU_\mu(R_1^2+R_2^2)}{U_4+ic}. \qquad (3.5)$$

The expressions (3.4) and (3.5) lead us to

$$\frac{c\overline{\psi^{class}}\gamma_\mu\psi^{class}}{\overline{\psi^{class}}\psi^{class}} = U_\mu. \qquad (3.6)$$

Let us evaluate (3.6) on the configurational coordinates $X_\mu$. We have:

$$\frac{dX_\mu}{d\tau} = \lim_{\hbar\to 0}\frac{dX_\mu}{d\sigma} = \lim_{\hbar\to 0}\frac{c\bar{\psi}\gamma_\mu\psi}{|\bar{\psi}\psi|} = \frac{c\overline{\psi^{class}}\gamma_\mu\psi^{class}}{\left|\overline{\psi^{class}}\psi^{class}\right|}.$$

Here $d\tau$ is the classical limit of $d\sigma$. In other words $\lim_{\hbar\to 0} d\sigma = d\tau$.

We can consider two cases:

Case $(i)$:

$$\frac{dX_\mu}{d\tau} = U_\mu = \frac{1}{m}\frac{\partial S}{\partial X_\mu} - \frac{e}{mc}A_\mu \qquad (3.7)$$

if $X_\mu$ is in a region where $\overline{\psi^{class}}\psi^{class} > 0$.

Case $(ii)$:

$$\frac{dX_\mu}{d\tau} = -U_\mu = -\frac{1}{m}\frac{\partial S}{\partial X_\mu} + \frac{e}{mc}A_\mu \qquad (3.8)$$

if $X_\mu$ is in a region where $\overline{\psi^{class}}\psi^{class} < 0$.

Case $(i)$ $\frac{dX_\mu}{d\tau} = U_\mu = \frac{1}{m}\frac{\partial S}{\partial X_\mu} - \frac{e}{mc}A_\mu$. Let us start writing (3.2) in the following way

$$\left(\frac{1}{m}\frac{\partial S}{\partial x_\mu} - \frac{e}{mc}A_\mu\right)\left(\frac{1}{m}\frac{\partial S}{\partial x_\mu} - \frac{e}{mc}A_\mu\right) = -c^2. \qquad (3.9)$$

If we evaluate (3.9) in the particle configurational coordinates $X_\mu$ we have $\frac{dX_\mu}{d\tau}\frac{dX_\mu}{d\tau} = -c^2$ or equivalently $dX_\mu dX_\mu = -c^2 d\tau^2$. The last expression reveals that $d\tau$ is precisely the particle proper time. Therefore *$d\sigma$ tends to the particle proper time when $\hbar \to 0$*. This result is very similar to the one obtained in [10] for a Bohmian Model built for a system of particles based on the Klein-Gordon equation [7]. If we derivate (3.9) with respect to the generic coordinate $x_\nu$ we obtain:

$$2\left(\frac{\partial S}{\partial x_\mu} - \frac{e}{c}A_\mu\right)\frac{\partial}{\partial x_\nu}\left(\frac{\partial S}{\partial x_\mu} - \frac{e}{c}A_\mu\right) = 0$$

$$\left(\frac{\partial S}{\partial x_\mu} - \frac{e}{c}A_\mu\right)\frac{\partial^2 S}{\partial x_\mu \partial x_\nu} = \frac{e}{c}\left(\frac{\partial S}{\partial x_\mu} - \frac{e}{c}A_\mu\right)\frac{\partial A_\mu}{\partial x_\nu}\ .$$

Then it is proven, following a simple algebraic procedure, that

$$\left(\frac{1}{m}\frac{\partial S}{\partial x_\mu}-\frac{e}{mc}A_\mu\right)\frac{\partial}{\partial x_\mu}\left(\frac{1}{m}\frac{\partial S}{\partial x_\nu}-\frac{e}{mc}A_\nu\right)=\frac{e}{mc}\left(\frac{1}{m}\frac{\partial S}{\partial x_\mu}-\frac{e}{mc}A_\mu\right)\left(\frac{\partial A_\mu}{\partial x_\nu}-\frac{\partial A_\nu}{\partial x_\mu}\right).$$

If we now evaluate this identity in the particle configurational coordinates $X_\mu$ we obtain that

$$\frac{d^2X_\nu}{d\tau^2}=U_\mu\frac{\partial}{\partial X_\mu}U_\nu=\frac{e}{mc}U_\mu F_{\mu\nu}$$

$$m\frac{d^2X_\nu}{d\tau^2}=\frac{e}{c}\frac{dX_\mu}{d\tau}F_{\mu\nu}\,.$$

Which constitues the Einsteinian equation for one particle ruled by the Lorentz force in Classical Relativistic Mechanics. That is, the ideas described by [9] produce the correct classical limit when we assume *the hypothesis of the phases equality on the classical limit.* The physical meaning of this hypothesis is that when we build the classical dynamics resulting from the limit when $\hbar\to 0$ on the original Bohmian system, the resulting physical system has a well defined action $S$.

Case $(ii)$ $\quad\frac{dX_\mu}{d\tau}=-U_\mu=-\frac{1}{m}\frac{\partial S}{\partial X_\mu}+\frac{e}{mc}A_\mu$

In this case, when we take the classical limit, following a completely analogous procedure we obtain

$$m\frac{d^2X_\nu}{d\tau^2}=-\frac{e}{c}V_\mu F_{\mu\nu}\,.$$

The essential difference is the change on the sign of the particle charge. With this ends the derivation of the classical limit for a Bohmian particle associated to the Dirac equation.

Summarizing, equation (3.2) along with equation (3.7) or (3.8) are the classical counterpart of the wave equation (2.1) and Eq. (2.5) (always assuming the hypothesis of the phases equality on the classical limit). In both cases we obtain adequate classical limits and the parameter $d\sigma$ tends to the proper time $d\tau$. The latter is entirely analogous to the case where the classical limit is taken on a Bohmian Model for the Klein-Gordon equation [7] in [10].

## 4. Two Particles.

For the sake of simplicity we will consider only two particles. The problem, however, can be extended to any finite number of particles $N$. This would be sort of an ideal gas where particles do not interact electromagnetically with each other (for some models in the bohmian literature [15-18]). In the model, however, the particles interact with an external electromagnetic field. The classical limit of a system with more than one particle will not be analyzed in this paper. We believe, however, that the hypothesis of the phases equality on the classical limit works for many particles as well.

The wave function for two particles has the following form:

$$\psi_{ij}\left(x_\mu^{(1)},x_\mu^{(2)}\right)\equiv\begin{pmatrix}\psi_{11}\left(x_\mu^{(1)},x_\mu^{(2)}\right) & \psi_{12}\left(x_\mu^{(1)},x_\mu^{(2)}\right) & \psi_{13}\left(x_\mu^{(1)},x_\mu^{(2)}\right) & \psi_{14}\left(x_\mu^{(1)},x_\mu^{(2)}\right)\\ \psi_{21}\left(x_\mu^{(1)},x_\mu^{(2)}\right) & \psi_{22}\left(x_\mu^{(1)},x_\mu^{(2)}\right) & \psi_{23}\left(x_\mu^{(1)},x_\mu^{(2)}\right) & \psi_{24}\left(x_\mu^{(1)},x_\mu^{(2)}\right)\\ \psi_{31}\left(x_\mu^{(1)},x_\mu^{(2)}\right) & \psi_{32}\left(x_\mu^{(1)},x_\mu^{(2)}\right) & \psi_{33}\left(x_\mu^{(1)},x_\mu^{(2)}\right) & \psi_{34}\left(x_\mu^{(1)},x_\mu^{(2)}\right)\\ \psi_{41}\left(x_\mu^{(1)},x_\mu^{(2)}\right) & \psi_{42}\left(x_\mu^{(1)},x_\mu^{(2)}\right) & \psi_{43}\left(x_\mu^{(1)},x_\mu^{(2)}\right) & \psi_{44}\left(x_\mu^{(1)},x_\mu^{(2)}\right)\end{pmatrix}.$$

The Pauli principle in terms of this expression would be

$$\psi_{ij}\left(x_\mu^{(2)},x_\mu^{(1)}\right)=-\psi_{ij}\left(x_\mu^{(1)},x_\mu^{(2)}\right). \qquad (4.1)$$

The quantity ${\psi_{il}}^*\beta_{lm}\beta_{ij}\psi_{jm}$ obtained by summing on repeated matrix indexes is a scalar and equal to

$$\begin{aligned}{\psi_{il}}^*\beta_{lm}\beta_{ij}\psi_{jm} &= |\psi_{11}|^2+|\psi_{12}|^2-|\psi_{13}|^2-|\psi_{14}|^2+|\psi_{21}|^2+|\psi_{22}|^2-|\psi_{23}|^2-|\psi_{24}|^2\\ &-|\psi_{31}|^2-|\psi_{32}|^2+|\psi_{33}|^2+|\psi_{34}|^2-|\psi_{41}|^2-|\psi_{42}|^2+|\psi_{43}|^2+|\psi_{44}|^2.\end{aligned}$$

This quantity is real. Every term of the expression is real. Of course this quantity could be negative in some domain of the space-time configurational space for the two particles.

We will try to extend Nikolic's idea. The wave equation would be formed by the following equations. In the first one the operator $p_\mu^{(1)}$ acts on the first particle generic coordinates $x_\mu^{(1)}$ and the operator $p_\mu^{(2)}$ on the second equation acts on the second particle generic coordinates $x_\mu^{(2)}$. $A_\mu^{(i)}$ is the electromagnetic field evaluated on the $i$-th particle generic coordinates; $i=1{,}2$.

$$\begin{aligned}\left(\gamma_\mu\right)_{ij}\left(p_\mu^{(1)}-\frac{e}{c}A_\mu^{(1)}\right)\psi_{jl} &= -mc\psi_{il}\\ \left(\gamma_\mu\right)_{ij}\left(p_\mu^{(2)}-\frac{e}{c}A_\mu^{(2)}\right)\psi_{lj} &= -mc\psi_{li.}\end{aligned}$$

We believe they are very similar to a model presented in [15] for many particles. On the other hand the Bohm-like guide equations would take the following form:

$$\begin{aligned}\frac{dX_\mu^{(1)}}{d\sigma} &= V_\mu^{(1)}=\frac{c{\psi_{il}}^*\beta_{lm}\left(\beta\gamma_\mu\right)_{ij}\psi_{jm}}{\left|{\psi_{il}}^*\beta_{lm}\beta_{ij}\psi_{jm}\right|}\\ \frac{dX_\mu^{(2)}}{d\sigma} &= V_\mu^{(2)}=\frac{c{\psi_{il}}^*\beta_{ij}\left(\beta\gamma_\mu\right)_{lm}\psi_{jm}}{\left|{\psi_{il}}^*\beta_{lm}\beta_{ij}\psi_{jm}\right|}.\end{aligned} \qquad (4.2)$$

It can be proved that both of these expressions determine four-vectors. On the right-hand side of the expressions (4.2) we sum over the repeated matrix indexes independently on the numerator and denominator. The right-hand side of both equations is evaluated on both particle configurational coordinates $\vec{X}^{(1)}$, $icT^{(1)}$, $\vec{X}^{(2)}$, $icT^{(2)}$. Here $d\sigma$ and $\left|{\psi_{il}}^*\beta_{lm}\beta_{ij}\psi_{jm}\right|$, once normalized, are respectively Nikolic time and Born-Nikolic probability distribution for this model.

The parameter $d\sigma$ is a scalar since everything else in (4.2) is covariant. To prove this (Appendix 2), they can be used the following facts:

*(i)* $\psi'_{sr} = S_{si}S_{rj}\psi_{ij}$ *that shows how the wave function transforms when passing from a frame of reference to another and* $S$ *is the transformation matrix that transforms one spinor from a frame of reference to another.*

*(ii)* $S^{+}\beta = \beta S^{-1}$. *This fact was already used on the first part when we proved that* $|\bar{\psi}\psi|$ *is a scalar.*

*(iii)* $\gamma_\mu = \frac{\partial x'_\eta}{\partial x_\mu} S^{-1}\gamma_\eta S$. *This relation is very important to prove that the Dirac equation for one electron has the same form in any inertial frame of reference.*

In [15] very interesting characteristics of this kind of models (Synchronized trajectories) are pointed out, even if the author perceives them as a flaw.

It may be interesting to do a little exercise in order to compare the model with Bell's reasoning [3] about one of his models. This reasoning was discussed on this paper introduction for the Non-Relativistic case. Let us consider a region of the space-time configuration $\overrightarrow{x_\mu^{(1)}}, ict^{(1)}, \overrightarrow{x_\mu^{(2)}}, ict^{(2)}$ that we will call $\Xi$ where $\psi_{il}{}^{*}\beta_{lm}\beta_{ij}\psi_{jm} > 0$. Then if the configurational coordinates $\vec{X}^{(1)}$, $icT^{(1)}$, $\vec{X}^{(2)}$, $icT^{(2)}$ are in this region $\Xi$ the equations (4.2) have the form:

$$\begin{aligned} \frac{dX_\mu^{(1)}}{d\sigma} &= \frac{c\psi_{il}{}^{*}\beta_{lm}\left(\beta\gamma_\mu\right)_{ij}\psi_{jm}}{\psi_{il}{}^{*}\beta_{lm}\beta_{ij}\psi_{jm}} \\ \frac{dX_\mu^{(2)}}{d\sigma} &= \frac{c\psi_{il}{}^{*}\beta_{ij}\left(\beta\gamma_\mu\right)_{lm}\psi_{jm}}{\psi_{il}{}^{*}\beta_{lm}\beta_{ij}\psi_{jm}}. \end{aligned} \tag{4.3}$$

These equations are very similar to the ones presented by J.S. Bell in [3] and in general represent a non-local model that, however, has the remarkable characteristic of being covariant. Let us assume now that the spinor is separable $\psi_{il} = \varphi_i\chi_l$ (here we are leaving aside the Pauli Exclusion Principle (4.1)). Then

$$\frac{dX_\mu^{(1)}}{d\sigma} = \frac{c\psi_{il}{}^{*}\beta_{lm}\left(\beta\gamma_\mu\right)_{ij}\psi_{jm}}{\psi_{il}{}^{*}\beta_{lm}\beta_{ij}\psi_{jm}} = \frac{c\varphi_i{}^{*}\chi_l{}^{*}\beta_{lm}\left(\beta\gamma_\mu\right)_{ij}\varphi_j{}^{*}\chi_m{}^{*}}{\varphi_i{}^{*}\chi_l{}^{*}\beta_{lm}\beta_{ij}\varphi_j\chi_m} = \frac{c\varphi_i{}^{*}\left(\beta\gamma_\mu\right)_{ij}\varphi_j{}^{*}}{\varphi_i{}^{*}\beta_{ij}\varphi_j}$$

and analogously

$$\frac{dX_\mu^{(2)}}{d\sigma} = \frac{c\chi_l{}^{*}\left(\beta\gamma_\mu\right)_{lm}\chi_m}{\chi_l{}^{*}\beta_{lm}\chi_m},$$

This kind of argument is used by Bell [3] to prove the difference between Bohmian dynamics when the particles are not separable and when they are. They illustrate the concept of non-locality in particles with spin. So we believe that with this kind of procedures it is possible in principle to extend much of Bell's ideas, about Bohmian dynamics, to Special Relativity.

The following continuity equations are a direct consequence of Dirac-like equations on which our model is based (Appendix 3):

$$\frac{\partial}{\partial x_\mu^{(1)}}\left(c\psi_{il}^*\beta_{lm}\left(\beta\gamma_\mu\right)_{ij}\psi_{jm}\right)=0$$
$$\frac{\partial}{\partial x_\mu^{(2)}}\left(c\psi_{il}^*\beta_{ij}\left(\beta\gamma_\mu\right)_{lm}\psi_{jm}\right)=0. \qquad (4.4)$$

Then

$$\frac{\partial}{\partial x_\mu^{(1)}}\left(\left|\psi_{il}^*\beta_{lm}\beta_{ij}\psi_{jm}\right|V_\mu^{(1)}\right)=0$$
$$\frac{\partial}{\partial x_\mu^{(2)}}\left(\left|\psi_{il}^*\beta_{lm}\beta_{ij}\psi_{jm}\right|V_\mu^{(2)}\right)=0. \qquad (4.5)$$

Let us add equations (4.5) and consider that $\rho\left(x_\mu^{(1)},x_\mu^{(2)}\right)=\left|\psi_{il}^*\beta_{lm}\beta_{ij}\psi_{jm}\right|$ does not depend on $\sigma$. We can write an entirely analogous equation to the one presented in [7, 10], that is, a relativistic conservation equation for the two particles system under consideration:

$$\frac{\partial\rho}{\partial\sigma}+\frac{\partial}{\partial x_\mu^{(1)}}\left[\rho V_\mu^{(1)}\right]+\frac{\partial}{\partial x_\mu^{(2)}}\left[\rho V_\mu^{(2)}\right]=0. \qquad (4.6)$$

Therefore if for $\sigma=0$ the particle configurational coordinates $X_\mu^{(1)}$, $X_\mu^{(2)}$ are distributed with probability distribution $\rho$ ("equivariance condition") these will remain so distributed for every $\sigma$. Now, it is satisfied the following equality for the jacobians

$$\frac{\partial\left(x^{(1)'},y^{(1)'},z^{(1)'},ct^{(1)'}\right)}{\partial(x^{(1)},y^{(1)},z^{(1)},ct^{(1)})}=\frac{\partial\left(x^{(2)'},y^{(2)'},z^{(2)'},ct^{(2)'}\right)}{\partial(x^{(2)},y^{(2)},z^{(2)},ct^{(2)})}=1$$

due to the Lorentz transformations [10]. The measure for an element with a fixed volume $dV$ on the configurational space, for two particles with generic coordinates $x_\mu^{(1)},x_\mu^{(2)}$, is independent on the frame of reference used for its calculus. Let us integrate $\rho$ over an arbitrary volumen of the two-particle configuration. Since the *element of the 8-dimensional volume is a scalar* and $\rho$ *is also a scalar*, the calculation of the probability of $X_\mu^{(1)},X_\mu^{(2)}$ being on $V$ at a given time $\sigma$ is independent on the frame of reference used for the calculation. If the two-particle configurational coordinates $X_\mu^{(1)}$, $X_\mu^{(2)}$ are distributed according to $\rho$ (once normalized) at time $\sigma=0$ *they will remain so for every* $\sigma$ *and for every inertial frame of reference.* We do not see any reason why these arguments could not be extended to a larger number of particles. In that case the spinor would have the form

$$\psi:\mathbb{R}^{4N}\quad\rightarrow\quad\mathbb{C}^{4^N}$$

where $N$ is the number of particles.

## 5. Conclusions

A Bohmian dynamics associated to the Dirac equation has been built. Initially, this has been done for one particle and subsequently for two particles. This process can be easily extended to any finite number of particles. In the construction of these dynamics Nikolic's idea of a space-time probability density (or in general a space-time configuration) has been used and $\sigma$ as a scalar parameter for the dynamics has been introduced. In the example for one particle the classical limit following the ideas of [9] has been built. This was done only for a certain kind of spinor, which has the characteristic that its component phases have the same limit when $\hbar \rightarrow 0$. We consider that this example is not only special *but very relevant* since an adequate classical limit was obtained. This problem needs further study. The example for two particles is interesting because it is non-local and covariant (which is a characteristic of the model built by Nikolic [7, 10] for a Bohmian Model based on the Klein-Gordon equation). Finally a very similar analysis to the one done by J.S. Bell for separable and non-separable Bohmian dynamics is obtained [3].

*Acknowledgements*

*I am grateful to Sheldon Goldstein for his invaluable help in my understanding of the ideas of his school and his important objections to reference* [10] *which I try to take into account here. I am grateful to Ernesto Hernández-Zapata for very useful discussions and for introducing me, some years ago, to the ideas of Louis de Broglie, David Bohm, J.S. Bell and the modern bohmian thought.*

**Appendix 1**

In the following steps we show a calculus to study the limit of *(i)* and *(iii)* (Section 3) under the Hipotesis of phases equality in the Classical Limit:

$$-\hbar^2\frac{\partial^2\psi}{\partial x_\mu\partial x_\mu}=-\hbar^2\frac{\partial}{\partial x_\mu}\begin{pmatrix}\frac{\partial R_1}{\partial x_\mu}e^{i\frac{S_1}{\hbar}}+\frac{i}{\hbar}R_1\frac{\partial S_1}{\partial x_\mu}e^{i\frac{S_1}{\hbar}}\\ \frac{\partial R_2}{\partial x_\mu}e^{i\frac{S_2}{\hbar}}+\frac{i}{\hbar}R_2\frac{\partial S_2}{\partial x_\mu}e^{i\frac{S_2}{\hbar}}\\ \frac{\partial R_3}{\partial x_\mu}e^{i\frac{S_3}{\hbar}}+\frac{i}{\hbar}R_3\frac{\partial S_3}{\partial x_\mu}e^{i\frac{S_3}{\hbar}}\\ \frac{\partial R_4}{\partial x_\mu}e^{i\frac{S_4}{\hbar}}+\frac{i}{\hbar}R_4\frac{\partial S_4}{\partial x_\mu}e^{i\frac{S_4}{\hbar}}\end{pmatrix}=$$

$$-\hbar^2\begin{pmatrix}\frac{\partial^2 R_1}{\partial x_\mu^2}e^{i\frac{S_1}{\hbar}}+\frac{i}{\hbar}\frac{\partial R_1}{\partial x_\mu}\frac{\partial S_1}{\partial x_\mu}e^{i\frac{S_1}{\hbar}}+\frac{i}{\hbar}\frac{\partial}{\partial x_\mu}\left(R_1\frac{\partial S_1}{\partial x_\mu}\right)e^{i\frac{S_1}{\hbar}}-\frac{1}{\hbar^2}R_1\frac{\partial S_1}{\partial x_\mu}\frac{\partial S_1}{\partial x_\mu}e^{i\frac{S_1}{\hbar}}\\ \frac{\partial^2 R_2}{\partial x_\mu^2}e^{i\frac{S_2}{\hbar}}+\frac{i}{\hbar}\frac{\partial R_2}{\partial x_\mu}\frac{\partial S_2}{\partial x_\mu}e^{i\frac{S_2}{\hbar}}+\frac{i}{\hbar}\frac{\partial}{\partial x_\mu}\left(R_2\frac{\partial S_2}{\partial x_\mu}\right)e^{i\frac{S_2}{\hbar}}-\frac{1}{\hbar^2}R_2\frac{\partial S_2}{\partial x_\mu}\frac{\partial S_2}{\partial x_\mu}e^{i\frac{S_2}{\hbar}}\\ \frac{\partial^2 R_3}{\partial x_\mu^2}e^{i\frac{S_3}{\hbar}}+\frac{i}{\hbar}\frac{\partial R_3}{\partial x_\mu}\frac{\partial S_3}{\partial x_\mu}e^{i\frac{S_3}{\hbar}}+\frac{i}{\hbar}\frac{\partial}{\partial x_\mu}\left(R_3\frac{\partial S_3}{\partial x_\mu}\right)e^{i\frac{S_3}{\hbar}}-\frac{1}{\hbar^2}R_3\frac{\partial S_3}{\partial x_\mu}\frac{\partial S_3}{\partial x_\mu}e^{i\frac{S_3}{\hbar}}\\ \frac{\partial^2 R_4}{\partial x_\mu^2}e^{i\frac{S_4}{\hbar}}+\frac{i}{\hbar}\frac{\partial R_4}{\partial x_\mu}\frac{\partial S_4}{\partial x_\mu}e^{i\frac{S_4}{\hbar}}+\frac{i}{\hbar}\frac{\partial}{\partial x_\mu}\left(R_4\frac{\partial S_4}{\partial x_\mu}\right)e^{i\frac{S_4}{\hbar}}-\frac{1}{\hbar^2}R_4\frac{\partial S_4}{\partial x_\mu}\frac{\partial S_4}{\partial x_\mu}e^{i\frac{S_4}{\hbar}}\end{pmatrix}$$

Therefore, the following limit is obtained:

$$\lim_{\hbar\to 0}\left(-\hbar^2e^{-i\frac{S}{\hbar}}\frac{\partial^2\psi}{\partial x_\mu\partial x_\mu}\right)=\frac{\partial S}{\partial x_\mu}\frac{\partial S}{\partial x_\mu}\psi^{class}$$

The limit (*iii)* will be considered in the following steps:

$$\lim_{\hbar\to 0}\left(-\frac{2e\hbar}{ic}e^{-i\frac{S}{\hbar}}A_\mu\frac{\partial\psi}{\partial x_\mu}\right)=\lim_{\hbar\to 0}\left(-\frac{2e\hbar}{ic}e^{-i\frac{S}{\hbar}}A_\mu\begin{pmatrix}\frac{\partial R_1}{\partial x_\mu}e^{i\frac{S_1}{\hbar}}+\frac{i}{\hbar}R_1\frac{\partial S_1}{\partial x_\mu}e^{i\frac{S_1}{\hbar}}\\ \frac{\partial R_2}{\partial x_\mu}e^{i\frac{S_2}{\hbar}}+\frac{i}{\hbar}R_2\frac{\partial S_2}{\partial x_\mu}e^{i\frac{S_2}{\hbar}}\\ \frac{\partial R_3}{\partial x_\mu}e^{i\frac{S_3}{\hbar}}+\frac{i}{\hbar}R_3\frac{\partial S_3}{\partial x_\mu}e^{i\frac{S_3}{\hbar}}\\ \frac{\partial R_4}{\partial x_\mu}e^{i\frac{S_4}{\hbar}}+\frac{i}{\hbar}R_4\frac{\partial S_4}{\partial x_\mu}e^{i\frac{S_4}{\hbar}}\end{pmatrix}\right)$$

$$= -\frac{2e}{c}\frac{\partial S}{\partial x_\mu} A_\mu \psi^{class}$$

The limits (*ii),* $(iv)$, *(v) y (vi)* can be established in a more direct way.

**Appendix 2**

It will be proved that the particles´s currents

$$\begin{aligned} J_\mu^{(1)} &= c\psi_{lm}^* \beta_{mj} \left(\beta\gamma_\mu\right)_{li} \psi_{ij} \\ J_\mu^{(2)} &= c\psi_{lm}^* \beta_{li} \left(\beta\gamma_\mu\right)_{mj} \psi_{ij} \end{aligned} \qquad (1)$$

are four-vectors (Section 4). The current for particle 1 in another frame of reference is :

$$J_\eta^{(1)'} = c\psi_{pq}^{'*} \beta_{qr} \left(\beta\gamma_\eta\right)_{ps} \psi_{sr}^{'}$$

$$= c\psi_{lm}^* S_{lp}^+ S_{mq}^+ \beta_{qr} \left(\beta\gamma_\eta\right)_{ps} S_{si} S_{rj} \psi_{ij} . \qquad (2)$$

To write (2) the following relationship has been used:

$$\psi_{pq}^{'} = S_{pl} S_{qm} \psi_{lm}$$

If this expression is conjugated $\psi_{pq}^{'*} = S_{pl}^* S_{qm}^* \psi_{lm}^* = \psi_{lm}^* S_{lp}^+ S_{mq}^+$ is obtained. This last equality has been used to in order to obtain (2). In the following step the expression (2) is rewritten using the relation $S^+\beta = \beta S^{-1}$:

$$J_\eta^{(1)'} = c\psi_{lm}^* S_{lp}^+ S_{mq}^+ \beta_{qr} \beta_{pn} \left(\gamma_\eta\right)_{ns} S_{si} S_{rj} \psi_{ij}$$

$$= c\psi_{lm}^* \left(S_{lp}^+ \beta_{pn}\right)\left(S_{mq}^+ \beta_{qr}\right)\left(\gamma_\eta\right)_{ns} S_{si} S_{rj} \psi_{ij}$$

$$= c\psi_{lm}^* \left(\beta_{lp} S_{pn}^{-1}\right)\left(\beta_{mq} S_{qr}^{-1}\right)\left(\gamma_\eta\right)_{ns} S_{si} S_{rj} \psi_{ij}$$

The following step is to change the order on the terms in this expression. Then, the relation $\gamma_\mu = \frac{\partial x_\eta^{'}}{\partial x_\mu} S^{-1} \gamma_\eta S$ is used (this equality is very important in proving that the Dirac equation has the same form in every frame of reference) :

$$J_\eta^{(1)'} = c\psi_{lm}^* \beta_{lp} \beta_{mq} \left(S_{pn}^{-1} \left(\gamma_\eta\right)_{ns} S_{si}\right) \left(S_{qr}^{-1} S_{rj}\right) \psi_{ij} . \qquad (3)$$

By multiplying (3) for $\frac{\partial x_\eta^{'}}{\partial x_\mu}$ the following identity is obtained:

$$\frac{\partial x_\eta^{'}}{\partial x_\mu} J_\eta^{(1)'} = c\psi_{lm}^* \beta_{lp} \beta_{mq} \left(\frac{\partial x_\eta^{'}}{\partial x_\mu} S_{pn}^{-1} \left(\gamma_\eta\right)_{ns} S_{si}\right) \delta_{qj} \psi_{ij}$$

$$= c\psi^*_{lm}\beta_{lp}\beta_{mj}\left(\gamma_\mu\right)_{pi}\psi_{ij}$$

$$= c\psi^*_{lm}\beta_{mj}\left(\beta\gamma_\mu\right)_{li}\psi_{ij} = J^{(1)}_\mu$$

Sumarizing:

$$J^{(1)}_\mu = \frac{\partial x'_\eta}{\partial x_\mu} J^{(1)'}_\eta$$

and analogously:

$$J^{(2)}_\mu = \frac{\partial x'_\eta}{\partial x_\mu} J^{(2)'}_\eta .$$

Therefore, the particles´s currents are four-vectors. Next, the expression $\psi^*_{il}\beta_{lm}\beta_{ij}\psi_{jm}$ is considered. It´ll be proved that is a scalar.

$$\psi'^*_{pq}\beta_{qr}\beta_{ps}\psi'_{sr} = \psi^*_{lm}S^+_{lp}S^+_{mq}\beta_{qr}\beta_{ps}S_{si}S_{rj}\psi_{ij}$$

$$= \psi^*_{lm}\left(S^+_{lp}\beta_{ps}\right)\left(S^+_{mq}\beta_{qr}\right)S_{si}S_{rj}\psi_{ij}$$

$$= \psi^*_{lm}\left(\beta_{lp}S^{-1}_{ps}\right)\left(\beta_{mq}S^{-1}_{qr}\right)S_{si}S_{rj}\psi_{ij}$$

$$= \psi^*_{lm}\beta_{lp}\beta_{mq}\left(S^{-1}_{ps}S_{si}\right)\left(S^{-1}_{qr}S_{rj}\right)\psi_{ij}$$

$$= \psi^*_{lm}\beta_{lp}\beta_{mq}\delta_{pi}\delta_{qj}\psi_{ij}$$

$$= \psi^*_{lm}\beta_{li}\beta_{mj}\psi_{ij}$$

Therefore it follows $\psi'^*_{pq}\beta_{qr}\beta_{ps}\psi'_{sr} = \psi^*_{lm}\beta_{li}\beta_{mj}\psi_{ij}$ and this expression is a scalar. The expression $\left|\psi^*_{lm}\beta_{li}\beta_{mj}\psi_{ij}\right|$, the candidate to be the probability density distribution in space-time configuration for the two-particle system, is also a scalar. Therefore, the bohmian velocities:

$$\frac{dX^{(1)}_\mu}{d\sigma} = \frac{c\psi^*_{lm}\beta_{mj}\left(\beta\gamma_\mu\right)_{li}\psi_{ij}}{\left|\psi^*_{lm}\beta_{li}\beta_{mj}\psi_{ij}\right|}$$
$$\frac{dX^{(2)}_\mu}{d\sigma} = \frac{c\psi^*_{lm}\beta_{li}\left(\beta\gamma_\mu\right)_{mj}\psi_{ij}}{\left|\psi^*_{lm}\beta_{li}\beta_{mj}\psi_{ij}\right|}$$

are four vectors.

**Appendix 3**

In the following expression let $l$ to be a fixed value chosen between 1, 2, 3 and 4:

$$\psi^*_{il}\left(\beta\gamma_\mu\right)_{ij}\psi_{jl} .$$

This expression is a function of the generic variables $x_\nu^{(1)}$ and $x_\nu^{(2)}$. For a while $x_\nu^{(2)}$ will be considered fixed. Once $l$ and $x_\nu^{(2)}$ are fixed, the expression $\psi_{jm\,.}$ will be a column of the spinor that satisfies the Dirac equation:

$$\left(\gamma_\mu\right)_{ij}\left(p_\mu^{(1)}-\frac{e}{c}A_\mu^{(1)}\right)\psi_{jl}=-mc\psi_{il}$$

Then, it can be concluded that the following continuity equation is true:

$$\frac{\partial}{\partial x_\mu^{(1)}}\left(\psi_{il}^*\left(\beta\gamma_\mu\right)_{ij}\psi_{jl}\right)=0. \qquad (1)$$

This is a very known consequence of the Dirac equation. Like a consequence of the nature of $\beta$:

$$\beta=\begin{pmatrix}1 & 0 & 0 & 0\\ 0 & 1 & 0 & 0\\ 0 & 0 & -1 & 0\\ 0 & 0 & 0 & -1\end{pmatrix}$$

The following expression can be developed like:

$$\psi_{im}^*\beta_{ml}\left(\beta\gamma_\mu\right)_{ij}\psi_{jl}=\psi_{i1}^*\left(\beta\gamma_\mu\right)_{ij}\psi_{j1}+\psi_{i2}^*\left(\beta\gamma_\mu\right)_{ij}\psi_{j2}-\psi_{i3}^*\left(\beta\gamma_\mu\right)_{ij}\psi_{j3}-\psi_{i4}^*\left(\beta\gamma_\mu\right)_{ij}\psi_{j4}.$$

Each term of the right hand in this identity satisfies the continuity equation (1). It follows immediately that:

$$\frac{\partial}{\partial x_\mu^{(1)}}\left(\psi_{im}^*\beta_{ml}\left(\beta\gamma_\mu\right)_{ij}\psi_{jl}\right)=0. \qquad (2)$$

This is in perfect accord with the definition of the particle current for particle 1. Analogously

$$\frac{\partial}{\partial x_\mu^{(2)}}\left(\psi_{im}^*\beta_{ij}\left(\beta\gamma_\mu\right)_{ml}\psi_{jl}\right)=0. \qquad (3)$$

These results are in agreement with the definition of particles' currents (Section 4).